\documentclass[useAMS,usegraphicx,usenatbib]{mn2e}

\title[Chemical evolution of the bulge of M31] {Chemical evolution of the bulge of M31: predictions about abundance ratios}
\author[Marcon-Uchida et al.]
{M. M. Marcon-Uchida $^1$ \thanks{E-mail: monica.uchida@cruzeirodosul.edu.br}, 
F. Matteucci $^{2,3,4}$ \thanks{E-mail: matteucci@oats.inaf.it}, 
G. A. Lanfranchi $^1$, 
E. Spitoni $^2$, \newauthor
V. Grieco$^{2,4}$\\
$^1$N\'ucleo de Astrof\'\i sica Te\'orica, Universidade
Cruzeiro do Sul, R. Galv\~ao Bueno 868, Liberdade, 01506-000, S\~ao Paulo, SP, Brazil\\
$^2$ Dipartimento di Fisica, Sezione di Astronomia, Universit\'a degli Studi di Trieste,
  Via G. B. Tiepolo 11, 34131 Trieste, Italy\\
$^3$ INAF Trieste, Via G. B. Tiepolo 11, 34131 Trieste, Italy\\
$^4$ INFN Trieste, Via A. Valerio, 2, 34134 Trieste, Italy\\
}

\usepackage[percent]{overpic}

\begin{document}
\date{Accepted . ; in original form xxxx}

\pagerange{\pageref{firstpage}--\pageref{lastpage}} \pubyear{xxxx}

\maketitle

\label{firstpage}

\begin{abstract}

We aim at reproducing the chemical evolution of the bulge of M31 by means of 
a detailed chemical evolution model, including radial gas flows coming from the disk. 
We study the impact of the initial mass function, the star formation rate and the time scale for bulge formation on 
the metallicity distribution function of stars.
We compute several models of chemical evolution using the metallicity distribution of dwarf stars as an observational constraint for the bulge of M31. Then, by means of the model which best reproduces the metallicity distribution function, we predict the [X/Fe] vs. [Fe/H] relations for several chemical elements (O, Mg, Si, Ca, C, N). 
Our best model for the bulge of M31 is obtained by means of a robust statistical method and assumes a Salpeter initial mass function, a Schmidt-Kennicutt law for star formation with an exponent k=1.5, an efficiency of star formation of $\sim 15\pm 0.27\, Gyr^{-1}$, and an infall timescale of $\sim 0.10\pm 0.03$Gyr. 
Our results suggest that the bulge of M31 formed very quickly by means of an intense star formation rate and an initial mass function flatter than in the solar vicinity but similar to that inferred for the Milky Way bulge.
The [$\alpha$/Fe] ratios in the stars of the bulge of M31 should be high for most of the [Fe/H] range, as is observed in the Milky Way bulge. These predictions await future data to be proven.

\end{abstract}


\section{Introduction}

The study of the chemical evolution of nearby galaxies is a very important tool to improve our knowledge about the process of star formation and galaxy evolution. The proximity of Local Group galaxies allows us to study the chemical properties of their stellar populations. 
The Andromeda galaxy (M31) is the largest and most massive galaxy of the Local Group of galaxies. Despite its
vicinity we have a small amount of information to constrain the chemical evolution
models, which is due to its high inclination angle $i=77^o$ (Walterbros \& Kennicutt 1987) which allows us only a near 
edge-on view of the system. In spite of that, individual stars were studied in the bulge of this galaxy and 
the stellar metallicty distribution could be inferred allowing one to investigate its formation.

Bulges are spheroidal stellar systems located in the center of spiral galaxies which usually
can be distinguished from the spiral disk by their different dynamics, chemistry and photometric
features. The so-called classical bulges present similar properties to elliptical galaxies and are 
mainly composed by an old stellar population, since this type of bulge is supposed to have formed very quickly at the beginning
of the galaxy evolution. 
Elmegreen (1999) suggested that in a classical bulge the potential well is too deep to have allowed a self-regulation
mechanism or gas outflows as it occurs in spiral disks and dwarf galaxies. Therefore, classical bulges should have
passed through a very intense phase of star formation, during a short timescale.

The bulge of M31 is a clear example of classical bulge (Kormendy \& Kennicutt 2004) and is one of the
few bulges for which we could detect individual stars and objects (Sarajedini \& Jablonka 2005, hereafter SJ05,
Worthey et al. 2005) that can constrain the chemical evolution models. 
More recent studies considering Lick indices of M31 bulge stars (Saglia et al. 2010), concluded that with the exception of the region in the inner arcsecs, the stars are uniformly old ($\ge$ 12 Gyr) and with an overabundance of $\alpha$-elements [$\alpha$/Fe]$\sim$ +0.2 dex and solar metallicity.

It is interesting to compare the properties of the Galactic bulge and the bulge of M31;
Matteucci \& Brocato (1990), hereafter MB90, first suggested that to reproduce the evolution of the
Milky Way bulge one should assume a fast gas collapse (timescales shorter than 1 Gyr), a very
efficient star formation rate (SFR) and an initial mass function (IMF) flatter than the 
Salpeter (1955) one. MB90 also predicted that the [$\alpha$/Fe] ratios in the Galactic bulge
should be supersolar over a large fraction of the metallicity range,
as a consequence of assuming a fast and intense bulge evolution, and this prediction was later on confirmed by observations (McWilliam \& Rich, 1994). More recently, detailed spectroscopic studies of the Galactioc bulge have confirmed these results (e.g. Johnson et al. 2011,2012.2013,2014; Bensby et al. 2013). In fact, a fast and intense evolution quickly produces a high Fe abundance as a consequence of the pollution  from the large number of core-collapse supernovae (SNe). Then, when the bulk of Fe is produced by Type Ia SNe, which explode with a time delay, the [Fe/H] in the interstellar medium (ISM) is already high. This is the reason for the large plateau predicted for the [$\alpha$/Fe] ratios.

Ballero et al. 2007a (hereafter BMOR2007) computed a chemical evolution model for the bulge of the Milky Way,
confirming that a short and intense period of star formation is needed to fit the observational constraints 
avaiable for the Galactic bulge at that time. This kind of model is typical of a star burst system with a very strong
SFR concentrated in the early evolutionary stages in analogy with elliptical galaxies. Following the same 
prescriptions, Ballero et al. (2007b, BKM2007) tested the universal IMF (Kroupa 2001) in the bulge of our
Galaxy and M31. The authors confirmed the result obtained by MB90 concluding that unlike disks for which one can use a
standard IMF (like the Salpeter one) in order to compute their chemical evolution, to reproduce the metallicity
distribution of both bulges an index of the IMF $x \sim 1$, flatter than the Salpeter (1955) index, is necessary. Finally, Cescutti \& Matteucci (2011) confirmed the results of BMOR2007 but found that even a Salpeter (1955) IMF can reproduce the more recent metallicity distribution function (MDF) of the Galactic bulge together with the [X/Fe] vs. [Fe/H] relations.

During the last few years the shape and the chemical evolution of the Milky Way bulge have been extensively discussed, unveiling a more complex scenario than simply the classical one.
With the advent of large  telescopes and large surveys new details about the kinematics, ages and chemical composition of individual stars in the Galactic
central region have been available in the literature (Zoccali et al. 2008; Babusiaux et al. 2010; Hill  et al. 2001; Gonzalez et al. 2011; Saito et al. 2011; Ness et al. 2013).
In particular, Hill et al. (2011) studied the chemical abundances in the center of the Galaxy and found two different stellar populations inside the bulge,
while  
Babusiaux et al. (2010) also found differences in the kinematics of these two populations, the metal poor being 
compatible with an old spheroid and  the metal rich being more compatible with a population formed by a secular
evolution process (pseudo-bulge).

In this paper,  we compute the chemical evolution of the bulge of M31 using a quite detailed adn updated version of the model built for the Milky Way bulge and adopted also for the M31 bulge  by BKM2007. This model includes gas infall and outflow, as in BKM2007, but it contains for the first time also radial gas flows from the disk.
We aim at reproducing the MDF and then at providing predictions for the [$\alpha$/Fe] ratios expected in the stars of M31 bulge. The model we adopt is predicting the evolution of several chemical species such as H, D, He, Li ,C, N,O, Ne, Mg, S, Si, Ca, Fe plus others. Here we will focus only on the evolution of H, C, N, O, Mg, Si, Ca and Fe.
The paper is organized as follows: in section 2 we present the observed MDF for the bulge of M31, in section 3 we briefly describe the 
adopted chemical evolution model,
in section 4 we present the model predictions and finally in section 5 we summarize our conclusions.

\section{The Metallicity Distribution in the bulge of M31}

The MDF of stars is one of the most important
observational constraints for chemical evolution models. It can provide us with fundamental  information about
the star formation and chemical enrichment histories.

SJ05 used Color Magnitude Diagrams (CMD) of stars in 
the bulge of M31 observed by the HST to derive the MDF for this system. Figure 1 shows the
observed distribution for 7771 stars as a function of [M/H], where M indicates a generic metallicity, namely the global metal content Z, which is dominated by oxygen.

\begin{figure} 
\begin{center}
\includegraphics[scale=0.40]{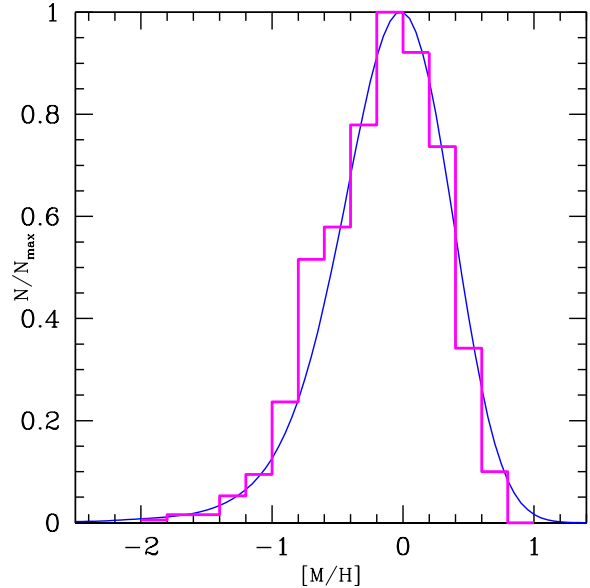}
\caption{Metallicity distribution for Sarajedini \& Jablonka (2005)(histogram) 
and the predictions of our best model (continuous line). Our predictions have been convolved with a gaussian with an error of 0.25 dex.}
\end{center}
\end{figure}

In SJ05 the authors also estimated  the effect of transforming their data as a function of [Fe/H], by adopting a relation between the [$\alpha$/Fe]= [M/H] ratio (where M is dominated by O) and the [Fe/H] ratio, as suggested by Ferraro et al. (2000). Unfortunately, this calibration relation assumes a fixed value for the [$\alpha$/Fe] ratio (+0.3 dex), and therefore it is an oversemplification since the [$\alpha$/Fe] ratio evolves with [Fe/H]. SJ05 concluded that the data as a function of [M/H] should then be scaled down by 0.2 dex. However, we think it is better to compare the theoretical distribution as a function of the [M/H] ratio with the data of SJ05. The quantity [M/H] is given by models of stellar evolution and includes $\alpha$-elements, where oxygen is the dominant element. 
On the other hand, BKM2007 compared the results of chemical evolution models with the SJ05 data as a function of [Fe/H], after applying the above transformation. They concluded that a good fit was obtained by assuming a fast bulge formation and a flat IMF, in agreement with the conclusions reached for the Milky Way bulge. In particular, they concluded that the IMF should be flatter than the universal IMF (UIMF) suggested by Kroupa (2001). Here we would like to test if the same conclusions hold when a fairer comparison of model predictions with data is performed.
In the SJ05 paper, the authors presented also the predictions of the closed-box model for chemical evolution adopting instantaneous recycling approximation and concluded that they obtained a good fit of the MDF of M31 except for the fact that they predicted too many low metallicity stars, in analogy with the G-dwarf problem in the solar vicinity. 
It is worth noting that instantaneous recycling approximation (I.R.A.) can indeed be applied to oxygen which is produced by core-collapse SNe on short timescales but the SJ05 model was a closed-box one and this is the reason they found a sort of G-dwarf problem.

In our model we assume that the bulge of M31 forms by infall of gas, which avoids overproduction of low metallicity stars.

\section{Model Prescriptions}

To reproduce the chemical evolution of the M31 bulge we used an updated version of the model proposed in BKM2007. The new feature of this model is the presence of radial gas flows from the disk of M31.
The prescriptions can be summarized as follows:

The bulge is assumed to be formed by infall of primordial gas described by an exponential
law, where $\tau$ is the timescale for the gas accretion.
\begin{equation}\label{infall} 
\frac{d\Sigma_I(R,t)}{dt}=A(R) e^{-\frac{t}{\tau}}, 
\end{equation} 
and $A(R)$ is a parameter obtained by reproducing the present time total surface mass density of M31's bulge (Geehan et al. 2006), and $\tau$ is the time scale for gas accretion.
The gas present in the bulge is supposed to be well mixed and homogeneous at any time.

We assumed that the star formation rate (SFR) is described by a simple power law (Schmidt 1959; Kennicutt, 1989) regulated by an 
efficiency $\nu$:
\begin{equation}\label{sfr}
 \Psi(R,t) \propto \nu\Sigma_g^k(R,t)
\end{equation}
where $\Sigma_g(r,t)$ is the gas surface mass density and $k$ is the slope of the Schmidt law for which we used 
two different values $k=1.0$ and $k=1.5$. As already demonstrated by other authors (BKMOR2007, 
MB90,) a high value for the efficiency in the SFR is needed to reproduce the chemical
evolution of the Galactic bulge. In this work we start by adopting a value of $\nu \sim 20 Gyr^{-1}$, as suggested by results for the MW bulge (e.g. MB90,  BMOR2007; Cescutti and Matteucci, 2011). 
We do not invoke a threshold in the star formation process, as was originally proposed for self-regulated disks, since there
is no evidence that it should exist also in bulges (Elmegreen 1999). However
we checked the model using a threshold of $4 M_\odot/pc^{-2}$ and no changes were observed in the results.

We included galactic wind in the model, which induces the gas to be removed from the star formation region but not necessarily to flow out of the Galactic potential well. The wind is treated as in BKM2007 and it develops when the thermal energy of gas overcomes the potential energy of gas. We find that the wind in M31 develops only after the bulk of stars has already formed (roughly at 1 Gyr from the beginning of the evolution). Therefore, the wind has a negligible effect on the chemical abundances but it regulates the star formation process, in the sense that very few stars are formed after 1 Gyr.

The novelty of this model is the inclusion of radial gas flows coming from the disk and we follow the prescriptions adopted in Spitoni et al. (2013), who developed a model for the disk of M31 with radial gas flows and galactic fountains, and where we direct the reader for details.
In particular, we adopted the following law for the radial gas flows in M31:
\begin{equation}
v_R=0.05R +0.45,
\end{equation}
where R represents the galactocentric distance. This law provides the best fit to the properties of the M31 disk. Here we assumed that the radial gas flows are only entering into the bulge from the disk.

The I.R.A. is relaxed, since stellar lifetimes are taken into account following
the prescriptions of Kodama (1997). The stellar yields used in this work are those of Woosley \& Weaver (1995)
corrected for metallicity dependent oxygen yields as suggested by Fran\c{c}ois et al. (2004). The rate for 
Type Ia SN is computed according to the work of Matteucci \& Recchi (2001) following the single degenerate scenario
for the progenitors of these supernovae.

In order to make the model as simple as possible we invoked a single slope IMF:
\begin{equation}\label{imf}
 \phi(M) \propto M^{-(1+x)}
\end{equation}
Many works have been dedicated to the study of the Initial Mass Function (IMF) 
and
its slope and universality are still subject of discussion. In our models we are going
to test the classical Salpeter IMF ($x=1.35$) and the MB90 IMF ($x=1.1$). This last IMF was favoured by BKM2007.

The assumed solar abundances adopted for normalizing the chemical abundances are those derived by Asplund et al. (2009).

The intrinsic parameters for M31 adopted in this work are bulge mass $M_{BM31}=3.3\times10^{10}M_\odot$ 
and bulge radius $R_{BM31}=4.0$ kpc (Geehan et al. 2006). 

\section{Results}

In this section we present the predictions for the chemical evolution of the classical bulge of M31.
As a first step we computed the model testing different scenarios for the M31 bulge evolution, by varying the main free parameters, especially  the IMF and the
infall timescale $\tau$.
BKM2007 also tested some models for the bulge of M31 reproducing the metallicity distribution of 
dwarf stars by SJ05. They concluded that the changes in the efficiency of the SFR and in $\tau$, the timescale of
gas accretion,
do not affect the position of the peak in the distribution but rather 
its width. The position of the peak is instead influenced mostly by the assumed  IMF. 
To reproduce the distribution observed by SJ05 we tested two different values for the slope of the IMF, two values for the exponent
 in the surface gas density in the Schmidt-Kennicutt law ($k=1.0$ and $k=1.5$) and two different timescales for the collapsing gas in 
the infall law ($\tau=0.1$ Gyr as proposed by BMOR2007 for the Milky Way galaxy and $\tau=0.05$ Gyr, to simulate a faster evolution).

\subsection{Metallicity Distribution}
Figure 1 shows the predicted metallicity distribution function (MDF) as a function of [M/H] compared to observational data.
After testing the various predicted MDFS, we found that the models with $k=1$ 
tend to produce more metal rich stars underestimating the number of metal poor objects
On the other hand, models with higher values of $k$ were tested and provided better results, in particular $k=1.5$, which agrees more with the Kennicutt (1998) suggested star formation rate. 
In order to test the validity of the selected free parameters ($\nu, \tau$) we have then made
use of a robust statistical method known as {\emph cross-entropy} 
(Rubinstein 1999)
to optimize the determination of the "best-fit" parameters to reproduce the observed MDF in the bulge of
M31. The cross-entropy has already been applied succesfully to solving astrophysical problems presented by Caproni
et. al. (2009) and references therein, demonstrating the great potential of this method.

Based on the best results that we obtained by using the Salpeter IMF (x=1.35) and exponent in the SFR equal to $k = 1.5$,  we fixed these parameters and used the cross entropy method to find the best
combination of the star formation efficiency ($\nu$) and time scale in the infall law ($\tau$) to reproduce
the MDF of SJ05. Figure 1 shows the best fit found by means of the 
statistical method, with the following values for the parameters, which are given with their theoretical error:
$\nu = 15.00 \pm 0.27 Gyrs^{-1}$ and $\tau = 0.10 \pm 0.03 Gyrs$.

\subsection{Abundance Ratios}

Chemical abundance ratios are a key parameter to understand the chemical evolution of stellar systems. 
In particular, the [$\alpha$/Fe] $vs$ [Fe/H] diagram clearly shows the effects of the time-delay between the chemical
enrichment from SNe Type II and SNe Type Ia. These ratios can, therefore, be used to constrain chemical evolution models and to provide hints concerning the evolution of the considered system.

To date, there are no available alpha-element abundances derived from observational data for the bulge of M31.
We present here the predictions for [O/Fe], [Mg/Fe], [Si/Fe], [Ca/Fe], [S/Fe], [C/Fe] and
[N/Fe] $vs$ [Fe/H] for the best model derived above, which best fits the metallicity distribution observed in the bulge of M31. 

\begin{figure} 
\includegraphics[width=0.45\textwidth]{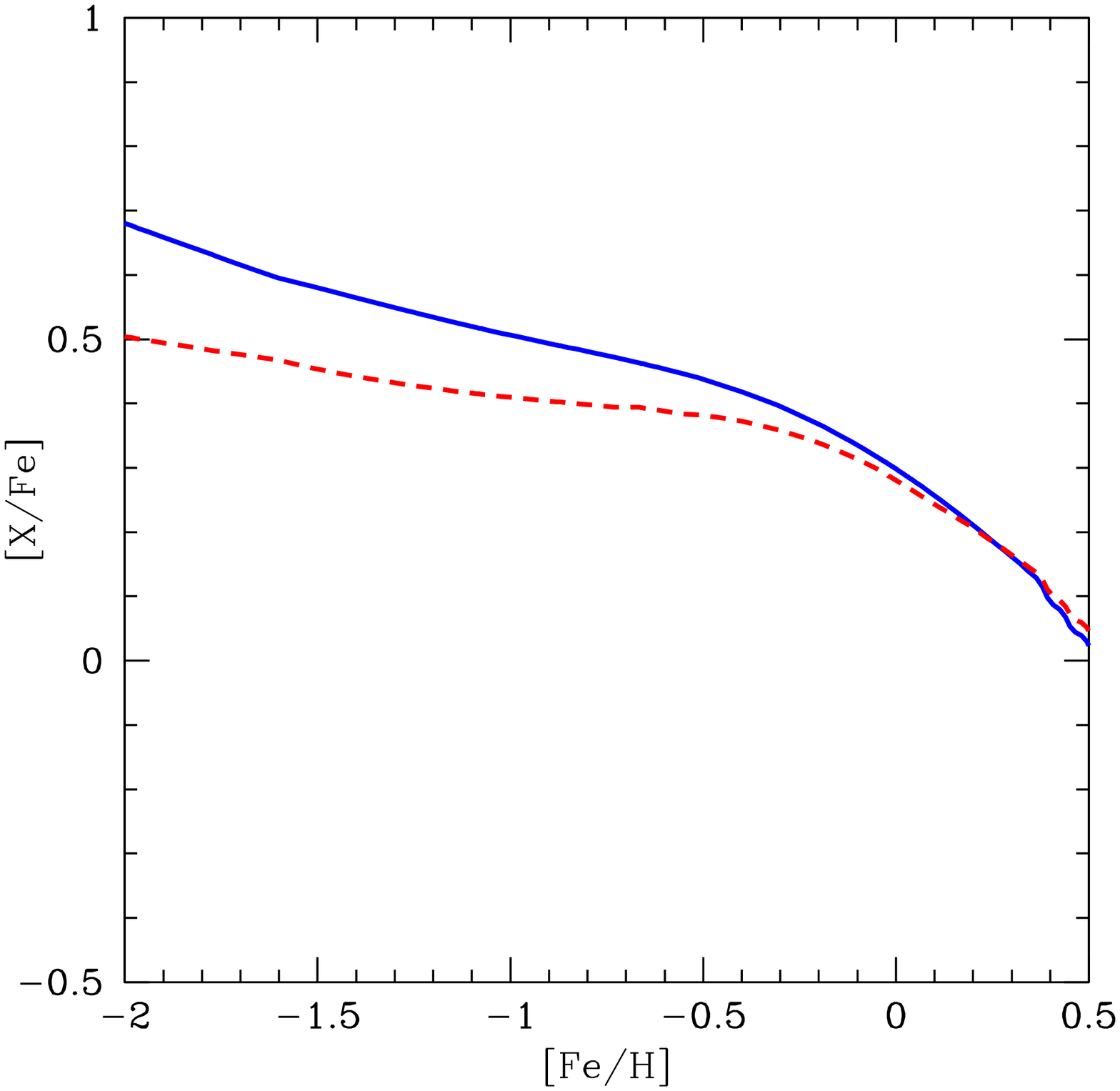} 
\includegraphics[width=0.45\textwidth]{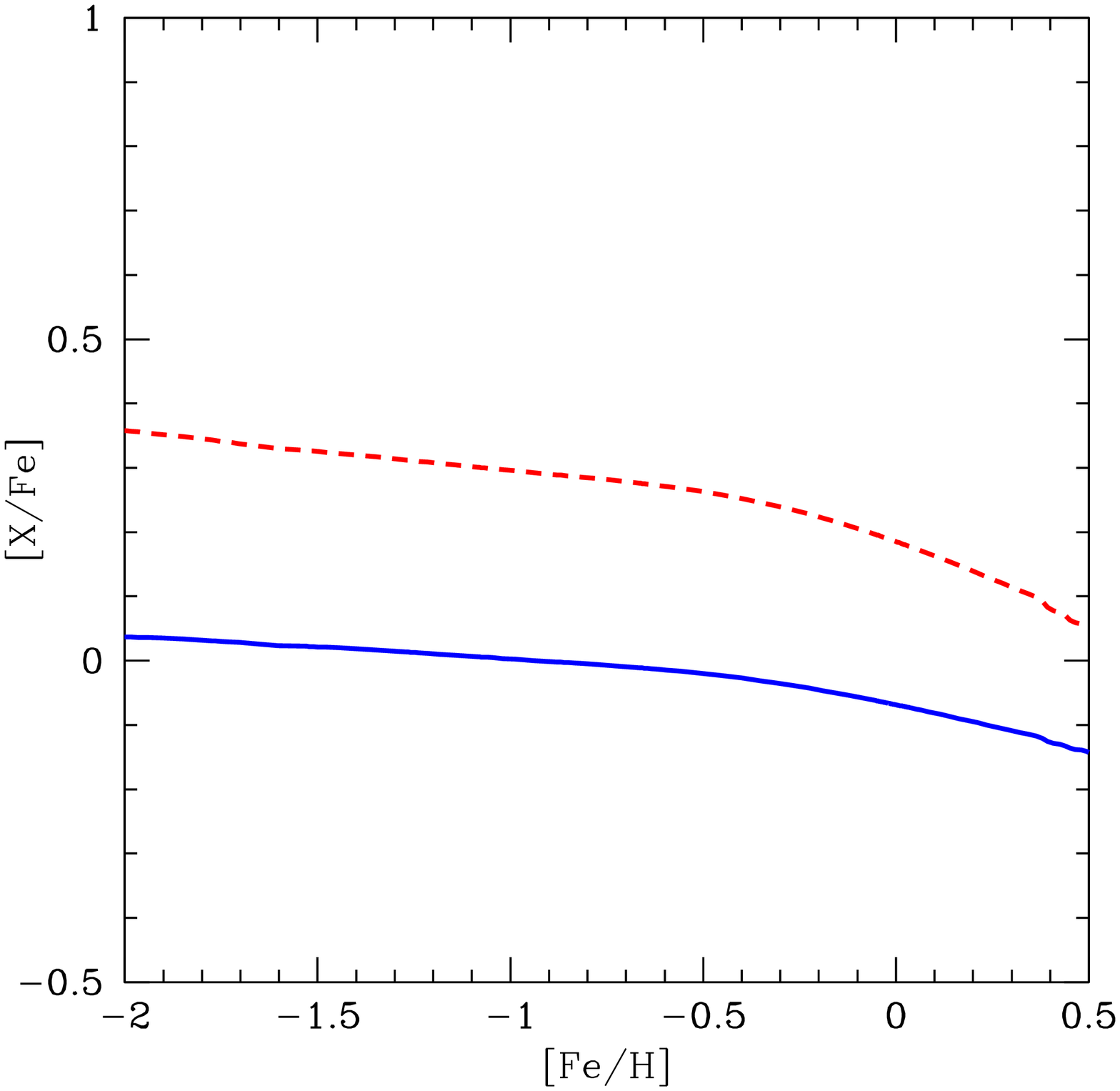} 
\caption{Predicted [$\alpha$/Fe] vs. [Fe/H] from our best model. In particular, in the top panel we show O (continuous blue line) and Mg (dashed red line), whereas in the bottom panel we show Si (dashed) and Ca (continuous).}
\end{figure}

\begin{figure} 
\includegraphics[width=0.45\textwidth]{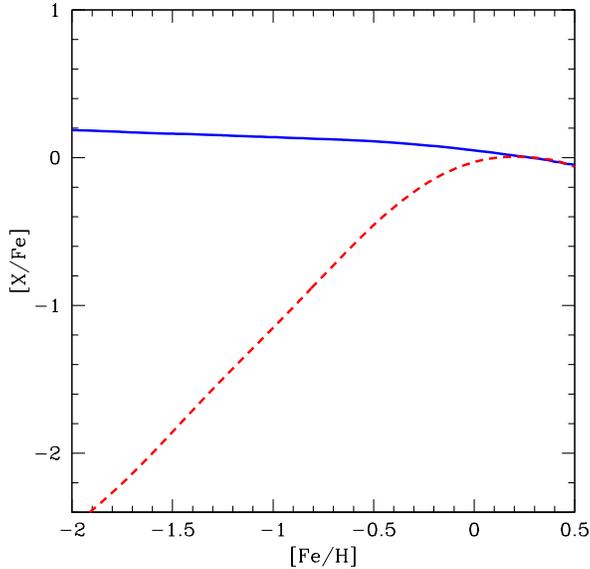} 
\caption{Predicted [N/Fe] and [C/Fe] vs. [Fe/H] from our best model. Nitrogen is the dashed red line, while C is the continuous blue one.}
\end{figure} 

The abundance ratios for the $\alpha$-elements O, Mg, Si, Ca  and S, as predicted by our model, present a 
similar trend, all of them with a long plateau characterized by oversolar abundance ratios. This behaviour is quite similar to what has been predicted for a classical Milky Way bulge (see for example Cescutti \& Matteucci, 2011 and Grieco et al. 2012) (see Section 4.3).
The long plateau present in all the plots of the  $alpha$-elements is due to the fact that the many Type II SNe (high star formation rate) produce a high [Fe/H] abundance in the gas before the first Type Ia SNe have time to start polluting the ISM.
As one can see, the overabundances of O and Mg are larger than those of Si and 
Ca, relative to Fe. The reason for that resides in the fact that Ca and Si are produced in a non-negligible way also by Type Ia SNe, whereas O and Mg are produced exclusively by core-collapse SNe. 
The [C/Fe] and [N/Fe] abundance ratios present an opposite trends, because the [C/Fe] relation slightly increases
with declining [Fe/H], whereas the [N/Fe] tends to decrease. The slight increase of C is due to the fact that C is produced in massive stars and also in low and intermediate mass stars. So, it shows a slight overabundance relative to Fe, similarly to what happens in the solar vicinity (see for example Romano et al. 2010). For N the situation is more complex because of the still existing uncertanties in its formation. The production of N in stars, in fact,  is still under discussion as it can result from 
both primary and secondary nucleosynthesis. In the first case nitrogen is created from H-burning by means of carbon which has been produced  by the
star, whereas in the secondary production N comes from the burning of C and O already present in the gas out of which the star formed. 
The secondary nucleosynthesis should be common to all stars while the primary production should occur in intermediate mass stars which go through dredge-up episodes during the asymptotic giant branch, as well as in massive rotating stars (see Meynet \& Maeder, 2002; Chiappini \& al. 2006). If primary N from massive stars of all metallicity is taken into account, as originally suggested by Matteucci (1986), the [N/Fe] ratio would be  almost constant all over the [Fe/H] range. In the Figure 3, [N/Fe] increases with [Fe/H] since here we have assumed that N is produced as a secondary element in massive stars. It is interesting to note that to explain N in the stars in the halo of the Milky Way, primary N from massive stars would be required (Chiappini et al. 2006). Thus, probably the same situation holds for the stars of M31's halo. 

\subsection{Comparison with the bulge of Milky Way}

Grieco et al. (2012) presented a chemical evolution model for the Milky Way bulge, taking into account the results of bulge surveys, modelling two different stellar populations: a metal-poor (MP) and a metal-rich (MR) one in the nuclear region of the Galaxy.
They concluded that these populations were formed in different episodes: the first one,
from which the MP population was produced, was described by typical classical bulge chemical evolution, whereas 
the second one, which produced the MR population, was assumed to form stars with a delayed time and out of pre-enriched gas with a longer
timescale for the infalling gas. 

The chemical evolution model for the MP of the Milky Way bulge presented by Grieco et al. (2012) is indeed very similiar to our classical bulge model for M31
predicting a very high efficiency star formation in a short time scale ($\nu_{MW} = \sim 25 Gyr^{-1}$ and $\tau_{MW}= 0.1$ Gyr). 
By comparing the two MDFs in Figure 4, one can notice that the MW bulge model shows a peak at a slightly higher metallicity and less metal poor stars than the bulge of M31, but they look substantially very similar, as should be expected from the assumptions made. Note that in Fig. 4 the two distributions are both computed as functions of [M/H], whereas often the MDF of the MW bulge as a function of [Fe/H] is compared to that of M31 as a function of [M/H]. The similarity between the two MDFs is not surprising since we assumed that both the MW bulge old population and the population of the bulge of M31 suffered strong bursts of star formation and evolved on quite short timescales. The small difference is then probably due to the different exponent for the gas density adopted in the SFR: $k=1$ for the MW bulge and $k=1.5$ for the M31 bulge.

Finally, in Fig. 5 we show a comparison between the predicted [Mg/Fe] vs. [Fe/H] for the bulge of M31 (best model) and for the Milky Way bulge from Grieco et al. (2012). We show Mg because is a typical $\alpha$-element measured in the Galactic bulge stars and the trends for the others $\alpha$'s are following the same behaviour of Mg. In particular, the MW bulge model predicts a  very similar  trend to that of M31 bulge. However, the [Mg/Fe] ratio declines slightly before and faster for the bulge of the Milky Way than for the bulge of M31; this is due to the more intense star formation in the MW bulge which consumes the gas faster and decreases before the star formation in the M31 bulge, with the consequence of having higher [Mg/Fe] ratios because the high Mg production related to the SFR, as compared to the Fe production from Type Ia SNe which  continues in spite of the low SFR.

\begin{figure}
\begin{center}
\includegraphics[scale=0.40]{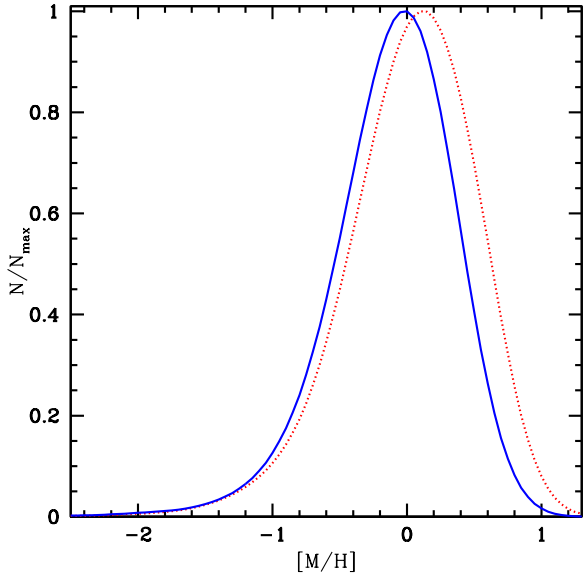}
\caption{Predicted MDFs for the MP population in the bulge of the Milky Way 
(Grieco et al. 2012) (dotted red line) and our
best model for the bulge of M31 (continuous blue line). Both distributions are functions of [M/H] and have been both convolved with a gaussian with an error of 0.25 dex.}
\end{center}
\end{figure} 

\begin{figure} 
\includegraphics[width=0.45\textwidth]{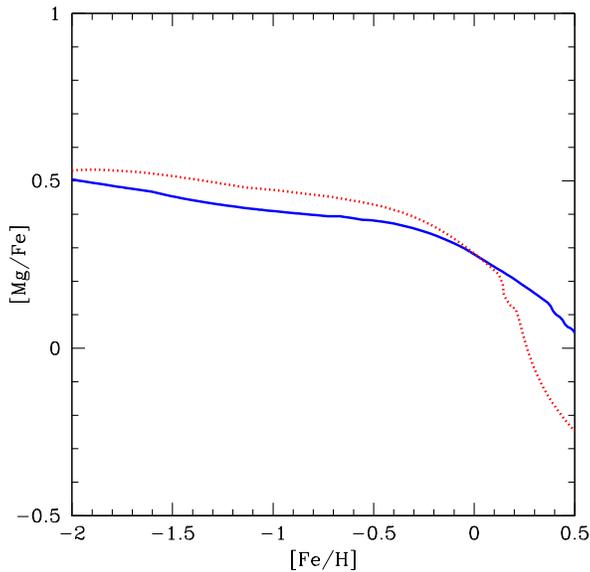}
\caption{Predicted [Mg/Fe] vs. [Fe/H] from our best model compared with the predictions for the MW bulge from Grieco et al. (2012).}
\end{figure}

\section{Conclusions}

In this work we reproduced the observed MDF of M31 by using an updated version of the chemical evolution
model for a classical bulge developed by BMOR2007 and BMK2007. 
The main differences between  the model of BKM2007 for the bulge of M31 and the present one, is our inclusion of radial gas flows entering the bulge from the disk and the adopted IMF. We have also predicted for the first time the chemical abundance ratios of $\alpha$-elements by means of the model that best fits the MDF of M31. These are real new predictions because there are not yet high resolution data relative to abundances in the stars in M31 bulge. 

Our conclusions can be summarized as follows:
\begin{itemize}

\item We have run several models testing the influence of some parameters 
such as the IMF, the
time scale for the infalling gas ($\tau$) and the exponent in the SFR ($k$).
We conclude that for the bulge of M31 (which is at least 1.5 times more
massive than the bulge of our Galaxy) a higher exponent for the gas surface mass density in the star formation law ($k=1.5$)
is required to reproduce the observed metallicity distribution. Models with $k=1.0$ produced more metal rich stars and underestimated the number of metal poor dwarfs in the bulge of M31. 
\item Previous papers (MB90 and BMOR2007) had concluded that to reproduce the chemical evolution of the bulge of the Milky
Way one has to use an IMF flatter than Salpeter one ($x \sim 1.1$), but Cescutti \& Matteucci (2011) suggested that also a Salpeter (x=1.35) IMF can reproduce the MW bulge MDF, as traced by the most recent data. 
In the present work we have shown that using a classical bulge
model with the Salpeter IMF we can reproduce the MDF of the bulge of M31, at variance with previous results (BKM2007). 
Moreover, we have used a robust statistical method (cross-entropy) to optimize the 'best-fit' for two free parameters of 
the model (the SFR efficiency and the infall timescale), taking the Salpeter IMF and the $k=1.5$ as fixed parameters.

\item By making use of the cross-entropy method to find the best fit for $\nu$ (efficiency of star formation) and $\tau$, the two main parameters of the model, 
we found our Best Model that suggests a quite high efficiency of star 
formation ($\nu =15.00  \pm 0.34 Gyr^{-1}$), and a short  timescale for the gas infall ($\tau =0.1 \pm 0.03 Gyr$), slightly lower than that suggested for the MW bulge (0.3-0.5Gyr, Cescutti \& Matteucci, 2011; Grieco et al. 2012). This model reproduces almost perfectly the observed MDF of SJ05. We point out that our 
theoretical MDF and the SJ05 one are both functions of the global metal content [M/H]. We have also tested a model without radial gas flows and the differences in the derived parameters are small. The small effect is due to the fact that the radial gas flows coming from the disk carry a modest amount of gas into the bulge, also because M31 bulge does not have a bar that could efficiently channel the gas via self-itersecting orbits.

\item The MDF distribution of the metal poor population in the bulge of the Milky Way (Grieco et al. 2012) is somewhat slightly more metal 
rich than the M31 bulge population but both systems  have experienced a similar evolution, except than in the Milky Way bulge there seems to be another stellar population formed as a consequence of secular evolution. The main differences among the predicted MDFs resides in the 
exponent assumed for the SFR which is $k=1.5$ for the M31 bulge and $k=1$ for 
the MW bulge, as well as the lack of radial gas flows in the MW bulge. It is worth recalling here that the main parameters influencing the MDF are the IMF, the efficiency of SF and the time-scale for gas infall. In particular, the IMF affects mainly the position of the peak in the MDF, while the efficiency of star formation plays a role in the shape of the MDF (the height of the peak and its extension in terms of [M/H]). The timescale of infall is not so important as the other two parameters but it influences the height of the peak (see BMOR2007). 

\item Concerning the $\alpha$-elements versus Fe,  we have obtained a long plateau and oversolar abundances for the [$\alpha$/Fe] $vs$ [Fe/H]
diagram for O, Mg, Si, S and Ca, similar to those observed for the Milky Way bulge (see Cescutti \& Matteucci, 2011 and references therein). There are also stars with lower [$\alpha$/Fe] ratios but they are a negligible fraction.  
Observational measurements of the 
abundances in the bulge of M31 are necessary to better constrain the 
chemical evolution models. Saglia et al. (2010) studied the bulge of M31 by means of Lick indices and simple stellar population models, and concluded that most of its stars are uniformely old ($\ge 12 Gyr$), of solar metallicity and showing moderate overabundances of $\alpha$-elements. We compared also the [Mg/Fe] vs.[Fe/H] predicted by Grieco et al. (2012) for the MP population of the Galactic bulge with the same relation predicted for M31. Even in this case, the two relations are very similar, thus confirming that the bulge of M31 and the MP population  of the Milky Way bulge formed very quickly by means of strong starburts, as indicated by the overabundances of $\alpha$-elements for a large range of [Fe/H] in both bulges. 
We hope that our results can be relevant also to a study of elliptical galaxies, given the similarity between classical bulges and elliptical galaxies of comparable mass (see Renzini, 2006 and references therein).
Future observations with very large telescopes will possibly be able to verify the validity of our predictions.

\end{itemize}

\section*{Acknowledgments}
M.M.Marcon-Uchida acknowledges important discussions with Luca Vincoletto.
M.M.Marcon-Uchida acknowledges financial support from the Brazilian agency 
FAPESP (process: 2010/17142-4). 
F.M., V.G. and E.S. acknowledge financial support from  PRIN MIUR-2010-2011, ``The Chemical and Dynamical Evolution of the Milky Way and Local Group Galaxies'', prot. N.2010LY5N2T. We thanks I.J. Danziger for reading the manuscript. We also thank the referee, R. M. Rich for important suggestions that improved the paper.

\end{document}